\documentclass[11pt]{article}
\usepackage{fa2023}
\usepackage{amsmath}
\usepackage{amsfonts}
\usepackage{cite}
\usepackage{graphicx}
\usepackage{color}
\usepackage{siunitx}
\usepackage{booktabs,makecell}
\usepackage{multirow}
\usepackage{layout}
\usepackage[utf8]{inputenc}
\usepackage[implicit=false]{hyperref}

\title{Differentiable Modelling of Percussive Audio with Transient and Spectral Synthesis}

\multauthor
{Jordie Shier$^{1*\dag}$ \hspace{1cm} Franco Caspe$^{1\dag}$ \hspace{1cm} Andrew Robertson$^2$} { \bfseries{Mark Sandler$^1$ \hspace{1cm} Charalampos Saitis$^1$ \hspace{1cm} Andrew McPherson$^3$}\\
  $^1$ Centre for Digital Music, Queen Mary University of London, United Kingdom\\
$^2$ Ableton AG, Berlin, Germany\\
$^3$ Dyson School of Design Engineering, Imperial College London, United Kingdom
\correspondingauthor{j.m.shier@qmul.ac.uk}{Shier et al.}
}

\begin{document}

\maketitle
\begin{abstract}
Differentiable digital signal processing (DDSP) techniques, including methods for audio synthesis, have gained attention in recent years and lend themselves to interpretability in the parameter space. However, current differentiable synthesis methods have not explicitly sought to model the transient portion of signals, which is important for percussive sounds. In this work, we present a unified synthesis framework aiming to address transient generation and percussive synthesis within a DDSP framework. To this end, we propose a model for percussive synthesis that builds on sinusoidal modeling synthesis and incorporates a modulated temporal convolutional network for transient generation. We use a modified sinusoidal peak picking algorithm to generate time-varying non-harmonic sinusoids and pair it with differentiable noise and transient encoders that are jointly trained to reconstruct drumset sounds. We compute a set of reconstruction metrics using a large dataset of acoustic and electronic percussion samples that show that our method leads to improved onset signal reconstruction for membranophone percussion instruments.

\end{abstract}
\keywords{\textit{Drum Synthesis, Differentiable DSP, Neural Networks}}

\section{Introduction}\label{sec:introduction}
Modeling of instrumental tones using data-driven methods and neural networks has received considerable attention in recent years. Differentiable digital signal processing (DDSP) enables classical audio synthesis algorithms to be integrated with neural networks and incorporated within gradient descent training regimes. DDSP synthesis methods have enabled high-quality and controllable audio synthesis with less training data compared to other deep learning approaches by leveraging the signal generation capabilities of DSP \cite{renault_differentiable_2022}. Additionally, these techniques have enabled rich synthesizer interaction methods, including audio-driven control of synthesizers, interpretable controls based on time-varying pitch envelopes, and timbre transfer applications.

The first work on DDSP by Engel et al. \cite{engel_ddsp_2020} used a differentiable sinusoidal plus noise synthesizer based on Serra's original method \cite{serra_spectral_1990} for instrumental audio synthesis. While a large body of research has followed, it has focused almost exclusively on modeling of harmonic tones with little application to unpitched percussion sounds. Furthermore, DDSP synthesis approaches have not sought to explicitly address the transient / onset portion of instrument signals, a signal component that is both important to musical audio perception \cite{siedenburg_specifying_2019} and is known to be poorly handled by sinusoidal plus noise synthesis \cite{verma_extending_2000}.
To this end, we explore sinusoidal plus noise modeling synthesis for non-harmonic percussion signals (i.e., drums and cymbals) and propose the addition of a temporal convolutional network (TCN) to handle transient / onset signal components in a differentiable framework.  Within our approach, a non-harmonic sinusoidal signal is generated using sinusoidal modeling synthesis (SMS) \cite{serra_spectral_1990} and parameter encoders for noise and transient synthesizers are jointly trained. A latent representation of the transient signal is generated by the transient encoder, paving the way for future work on controllable drum synthesis within a differentiable, data-driven paradigm.

In this work, we focus on the reconstruction of unpitched drum sounds to explore the strengths and weaknesses of our proposed method for drum synthesis. We train our model using a diverse dataset of electronic and acoustic drumset sounds and evaluate the results using a number of reconstruction metrics. We find that our proposed method improves onset reconstruction for membranophones, although adds artifacts that degrade the decay portion of the sound.
We conclude with suggestions for future work focused on controllable differentiable drum synthesis, which builds on the methods proposed here.

\section{Background}
\subsection{Drum Synthesis}
Synthesis of drum sounds has been explored extensively in previous work. In contrast to physically-informed approaches for percussion synthesis \cite{cook_physically_1997, bilbao_modular_2009}, spectral synthesis methods, which include sinusoidal modeling synthesis (SMS) \cite{serra_spectral_1990}, seek to model perceptual qualities of audio signals. SMS has been used for percussive synthesis \cite{smith_parshl_1987}, including as an analysis step for modal synthesis \cite{cook_physically_1997}. Recent research used a time-frequency analysis method and sinusoidal synthesis to model tom drum sounds that were indistinguishable from real sounds in a listening test \cite{kirby_evolution_2021}, pointing to the potential for SMS for high-quality, real-time drum synthesis. In contrast to traditional synthesis methods, data-driven approaches learn synthesis functions from a corpus of audio. Related work specific to drum synthesis has utilized convolutional neural networks \cite{ramires_neural_2020}, generative adversarial networks \cite{nistal_drumgan_2020}, and diffusion networks \cite{rouard_crash_2021}. %

\subsection{Differentiable Digital Signal Processing}
Differentiable digital signal processing (DDSP) combines the strengths of traditional DSP approaches with data-driven approaches. A significant body of work has explored the application of DDSP for the synthesis of harmonic instrumental tones \cite{engel_ddsp_2020, caspe_ddx7_2022}, including pitched percussion (i.e., piano) \cite{renault_differentiable_2022}. Directly estimating the frequencies of oscillators using gradient descent is a challenging problem due to the oscillatory nature of loss surfaces produced by current audio reconstruction objectives \cite{turian_im_2020, hayes_sinusoidal_2023}. Most approaches rely on pitch estimation algorithms and harmonic oscillators to circumvent this problem. As a result, DDSP has seen limited application in the modeling of non-harmonic sounds, and therefore unpitched percussion instruments. The one exception is the recent work by Diaz et al. \cite{diaz_rigid-body_2023}, which implemented a differentiable modal resonator. Here, we identify the lack of transient modeling as another limitation of current DDSP methods for synthesizing percussive sounds, and propose the use of TCNs to generate these signal components.

\subsection{Transient Modeling}
Transient and onset signals, defined by abrupt changes in amplitude, phase, or frequency information \cite{thornburg_analysis_2003}, are not well represented by sinusoidal plus noise models, which is the basis of many DDSP approaches. While musical onsets and transient regions are an important perceptual component for all musical sounds \cite{siedenburg_specifying_2019}, they are particularly important for percussive audio, which often contains rapidly decaying signal components following an impulsive event.
Levine and Smith \cite{levine_sines_1998} suggest that sinusoidal, noise, and transient signal components be modeled separately and proposed a method for isolating transients. Similar approaches proposed extracting attack or transient signal components \cite{serra_spectral_1990}, applying modifications to the sinusoidal signal components, and then re-inserting transients directly to the output. Verma and Meng \cite{verma_extending_2000} 
suggest that sinusoidal modeling can also be used to model transient signals if conducted in the correct signal domain, and propose doing so in the discrete cosine transform (DCT) domain. %
An alternative approach proposed a source filter method \cite{thornburg_analysis_2003}, reframing the transient modeling formulation as transient $*$ sines + noise. 
Conceptually, our approach is similar to this source filter formulation; however, we use gradient descent to estimate parameters for a TCN which acts as a filter for transient generation. %

\subsection{Temporal Convolutional Networks}
Temporal convolutional networks (TCNs) use multiple layers of time domain convolutions followed by non-linear activation functions and have been successfully applied to many deep learning audio tasks including audio synthesis \cite{oord_parallel_2018} and audio effect modeling \cite{steinmetz_efficient_2022}. Dilated convolution kernels with exponentially increasing dilation rates enable larger receptive fields with relatively few layers, helping to address temporal signal dependencies. A dilated TCN was used by Wang et al. \cite{wang_neural_2019} in a source-filter based approach for voice synthesis. We explore a similar approach for transient synthesis using TCNs and a sinusoidal input signal.

\section{Method}
\begin{figure*}[ht]
\includegraphics[width=0.9\textwidth]{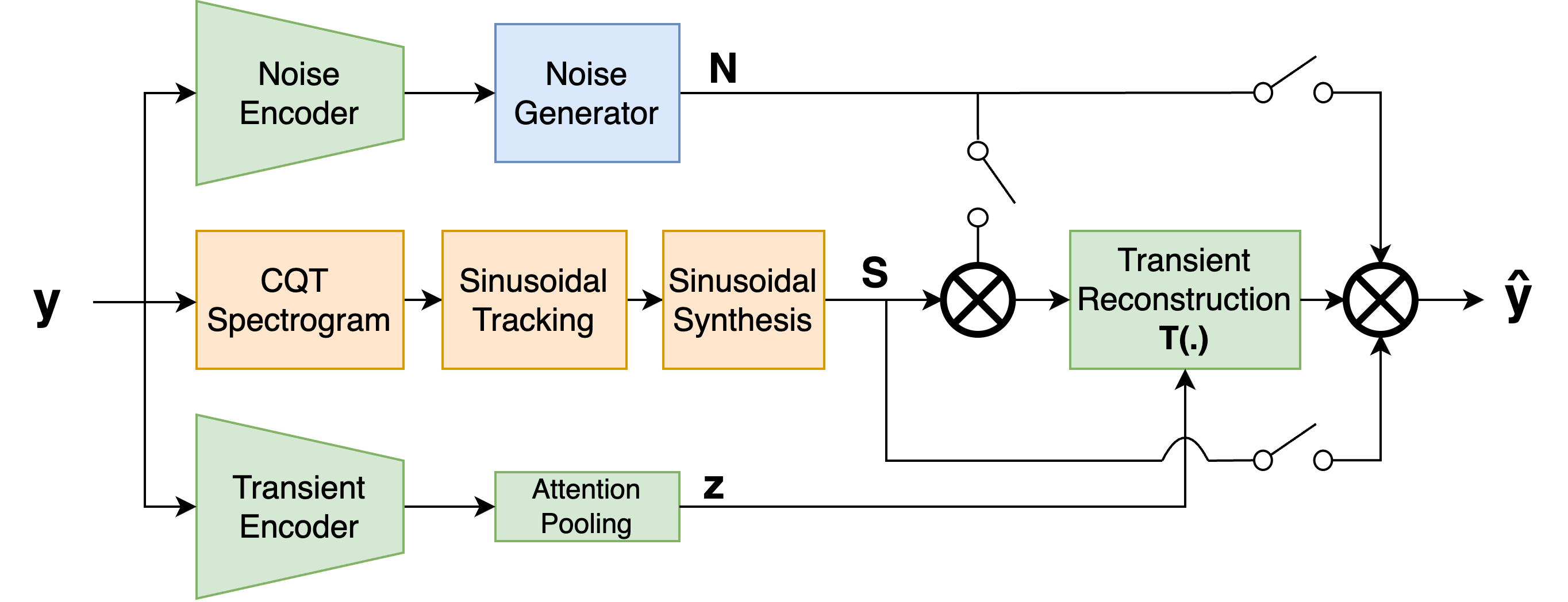}
\centering
\caption{
Our system models an input drum waveform $y$ using three parallel pipelines: sinusoids, noise, and transients.
The sinusoidal components are obtained by tracking, while noise and transient components are learned through differentiable modeling. A function, $T(.)$, is learned for reconstructing transients from the sinusoidal and noise components, which is controlled by a vector $z$ generated by a transient encoder.
We explore several strategies for mixing signal components, which are fed to $T(.)$ or directly to the output, as indicated by the switches in the figure. A mixing scheme is predefined for a model.
}
\label{fig:architecture}
\end{figure*}

Our proposed approach, shown in Figure \ref{fig:architecture}, consists of three parallel analysis and synthesis pipelines that aim to model an input drum signal $y$ using sinusoidal, noise, and transient components. Their outputs are then combined following 
a mixing strategy to resynthesize an approximation $\hat{y}$ of the input.

\subsection{Sinusoidal Modeling}
\label{sec:sinusoidal-modeling}
The sinusoidal pipeline seeks to model tonal signal components with a set of time-varying sinusoids.
In particular, the system aims to extract parameters $\langle a^j_i, f^j_i, \phi^j \rangle$, which denote the $i$th frame-wise amplitude and frequency, and the initial phase of the $j$th sinusoidal component, respectively. %
Due to the aforementioned challenges related to estimating frequency parameters using gradient descent, we instead employ a well-known tracking algorithm based on sinusoidal peak picking \cite{serra_spectral_1990}. We use the constant-Q transform (CQT) for our time-frequency representation as it provides improved frequency resolution, which is important for certain percussion sounds (e.g., kick drums). We track $N_S$ sinusoidal components on the CQT spectrogram of the input signal, with a minimum frequency of 20 Hz, 24 bins per octave, and a total frequency span of 10 octaves. Finally, we synthesize the sinusoidal components $S_n$ by linearly interpolating the amplitudes and frequencies from frame to sample rate and adding the extracted initial phase to each component, as shown in Equation \ref{eq:sinusoidals}.
\begin{equation}\label{eq:sinusoidals}
    S_n = \sum_{j=1}^{N_S}{a_n^j sin(2 \pi \cdot f_n^j \cdot n + \phi^j)}
\end{equation}

\subsection{Noise Modeling}
The noise pipeline consists of a frame-level analysis network implemented with a SoundStream encoder \cite{zeghidour_soundstream_2022} without its residual vector quantizer and a filtered noise generator with linear frequency bands based on \cite{engel_ddsp_2020}.
The encoder is composed of a 1D convolutional input layer, followed by a series of non-causal 1D convolutional blocks, each including three residual units of tripling dilation ratio, and a strided downsampling layer with ELU activation. Finally, an output convolutional layer projects the intermediate representation to the appropriate number of output channels.

The noise encoder takes the input audio and predicts a frame-wise sequence $\nu_i \in \mathbb{R}^{N_N}$ of gain coefficients for filters equally spaced on a linear frequency scale, with $i$ denoting the $i$th frame and $N_N$ the number of filter bands. 
Since our sinusoidal model presents high resolution on the low-frequency spectrum, we use a noise generator on a linear scale to account for the high frequencies of drum signals even when using a few noise bands.

Finally, the encoder's output is then relayed to the control input of the noise generator, which synthesizes an impulse response for each set of filter coefficients $\nu_i$ using the inverse discrete Fourier Transform, here denoted as $F^{-1}(.)$.
The inversion yields an impulse response of length $2N_N$ which is then padded with zeroes left and right up to a length $2h$. They are convolved with a white noise signal $\epsilon_n$ and then aggregated using overlap-add over $L$ Hanning windows $w(.)$ of length $2h$, with $h$ being the hop size of the operation.
The noise component $N$ is generated as shown in Equation \ref{eq:noise}.
\begin{equation}\label{eq:noise}
    N_n = \sum_{i=1}^{L}{w(n-hi)(\epsilon_n * F^{-1}(\nu_i))}
\end{equation}

\subsection{Transient Modeling}
Given a drum signal resynthesized with sinusoidal and noise components as input, we design a pipeline to generate transients that are not effectively modeled by the other components. The transient modeling pipeline has two goals: i) learn a controllable non-linear transfer function $T: x,z \rightarrow y$ that processes the sinusoidal and noise components as monophonic audio signals of length $L$, $x \in \mathbb{R}^L$ and generates an output with improved transients $y \in \mathbb{R}^L$
; ii) learn a controller network that predicts a transient conditioning vector $z \in \mathbb{R}^D$ to control transient synthesis for a particular input.

We implement the transfer function with a TCN architecture similar to \cite{steinmetz_efficient_2022}, which can be seen as a chain of non-linearities and filters that operate at the audio rate. 
We use Feature-wise Linear Modulation (FiLM) \cite{perez_film_2018} as a method to control the distortion and filtering characteristics of the TCN, conditioned on the encoded input drum signal $z$. FiLM functions by applying an affine transform to the output of the convolution of each layer of the TCN. A unique multi-layer perceptron is learned for each TCN layer and maps from the transient encoding $z$ produced by the controller network to shift and scale parameters for the affine transform. The controller network is implemented with a SoundStream encoder similar to the noise encoder. However, we use attention pooling to aggregate the frame-wise features predicted by the transient encoder into a single latent embedding vector $z$ used for FiLM.

\section{Experiments}
\sisetup{detect-weight=true,detect-inline-weight=math}
\begin{table*}[ht]
\small
\begin{center}
\setlength\tabcolsep{4pt}
\begin{tabular}{lSSSSSSSSS}
\toprule
\multicolumn{1}{c}{} & \multicolumn{3}{c}{All} & \multicolumn{3}{c}{Acoustic} & \multicolumn{3}{c}{Electronic} \\
\cmidrule(lr){2-4}
\cmidrule(lr){5-7}
\cmidrule(lr){8-10}
\thead{Method} & {\it MSS $\downarrow$} & {\it LSD $\downarrow$} & {\it SF $\downarrow$} & {\it MSS $\downarrow$} & {\it LSD $\downarrow$} & {\it SF $\downarrow$} & {\it MSS $\downarrow$} & {\it LSD $\downarrow$} & {\it SF $\downarrow$}\\
\midrule
$S$ & 1.34 & 2.51 & 153.7 & 1.66 & 3.64 & 80.9 & 1.03 & 1.38 & 226.6 \\
$S+N$ & 0.92 & \bfseries 1.20 & 152.6 & 1.06 & \bfseries 1.65 & 79.6 & 0.77 & \bfseries 0.75 & 225.8  \\
\midrule
$T(S)$ & 0.99 & 1.70 & 131.1 & 1.16 & 2.31 & 54.9 & 0.81 & 1.08 & 207.5  \\
$T(S + N)$ & 0.87 & 1.45 & 128.7 & 1.00 & 1.80 & 59.2 & 0.74 & 1.09 & 198.3  \\
$T(S) + N$ & \bfseries 0.85 & 1.36 & 128.8 & \bfseries 0.99 & 1.74 & \bfseries 54.6 & \bfseries 0.71 & 0.99 & 203.1 \\
$T(S) + S + N$ & 0.86 & 1.39 & \bfseries 123.4 & 1.00 & 1.75 & 56.5 & 0.73 & 1.02 & \bfseries 190.1 \\
\bottomrule
\end{tabular}
\caption{\small Overall reconstruction error metrics. Under method, $S$, $N$, and $T()$ refer to the sinusoidal, noise, and transient synthesis methods, respectively. MSS is multi-resolution spectral loss (the train loss), LSD is log spectral distance, and SF is spectral flux onset error.
}
\label{tab:overall-metrics}
\end{center}
\end{table*}

\sisetup{detect-weight=true,detect-inline-weight=math}
\begin{table*}[ht]
\small
\begin{center}
\setlength\tabcolsep{2.5pt}
\begin{tabular}{lSSSSSSSSSSSS}
\toprule
\multicolumn{1}{c}{} & \multicolumn{3}{c}{Kick} & \multicolumn{3}{c}{Snare} & \multicolumn{3}{c}{Tom} & \multicolumn{3}{c}{Cymbals}\\
\cmidrule(lr){2-4}
\cmidrule(lr){5-7}
\cmidrule(lr){8-10}
\cmidrule(lr){11-13}
\thead{Method} & {\it MSS $\downarrow$} & {\it LSD $\downarrow$} & {\it SF $\downarrow$} & {\it MSS $\downarrow$} & {\it LSD $\downarrow$} & {\it SF $\downarrow$} & {\it MSS $\downarrow$} & {\it LSD $\downarrow$} & {\it SF $\downarrow$} & {\it MSS $\downarrow$} & {\it LSD $\downarrow$} & {\it SF $\downarrow$}  \\
\midrule
$S$ & 1.22 & 2.18 & 470.5 & 1.37& 2.44 & 84.8 & 1.31 & 3.10 & 185.1 & 1.47 & 2.64 & 12.8 \\
$S+N$ & 0.92 & \bfseries 1.06 & 470.4 & 0.91 & \bfseries 1.03 & 84.0 & 0.81 & \bfseries 1.27 & 
185.0 & \bfseries 0.98 & \bfseries 1.43 & \bfseries 10.3 \\
\midrule
$T(S)$ & 0.93 & 1.60 & 388.8 & 0.93 & 1.47 & 62.8 & 0.93 & 1.83 & 181.2 & 1.11 & 1.90 & 11.7 \\
$T(S + N)$ & 0.79 & 1.34 & \bfseries 364.6 & 0.83 & 1.30 & 58.3 & 0.77 & 1.47 & 200.1 & 1.00 & 1.62 & 12.7 \\
$T(S) + N$ & \bfseries 0.76 & 1.28 & 379.6 & \bfseries 0.81 & 1.21 & 58.3 & \bfseries 0.75 & 1.41 & 183.0 & 0.99 & 1.53 & 12.6 \\
$T(S) + S + N$ & 0.77 & 1.30 & 372.1 & 0.84 & 1.24 & \bfseries 54.4 & 0.76 & 1.43 & \bfseries 167.0 & 1.01 & 1.56 & 11.4 \\
\bottomrule
\end{tabular}
\caption{\small Reconstruction metrics calculated on specific percussion instruments. Cymbals include hihats.}
\label{tab:instrument-metrics}
\end{center}
\end{table*}

We conducted a series of experiments to evaluate our approach in terms of audio reconstruction. To this end, we composed several different configurations of our model to help us understand the potential benefits of using a TCN to generate transient signal components. We consider four different mixing strategies: 1) transient generation from sines only, denoted $T(S)$; 2) transient generation from mixed sines and noise, denoted $T(S + N)$; 3) transient generation from sines, with noise added in parallel, denoted $T(S) + N$; 4) transient generation from sines, with noise and sines added in parallel, denoted $T(S) + S + N$. We baseline these approaches against sinusoids only $S$ and sinusoids plus noise only $S + N$. Audio examples of results and architecture details are provided on an accompanying website.\footnote{\url{https://jordieshier.com/projects/differentiable_transient_synthesis/}}

Due to the lack of high-quality, open-source one-shot drum samples, we curated our experimental dataset from the author's collection of commercial sample packs. While the FreeSound One-Shot Percussion dataset \cite{ramires_neural_2020} contains over 10k sounds, the dataset sample rate is 16kHz and samples aren't annotated with sound source or instrument. We used samples that were professionally produced at full audio resolution, allowing us to conduct our experiments at a sample rate of 48kHz, which we feel is important for evaluating transient signal components and improves the applicability of our method to music production contexts.

The final dataset contained 25k samples with an equal split of acoustic and electronic sources, and included kick, snare, tom, hihat, cymbals, and a variety of other percussion instruments. All samples were preprocessed to remove starting silence and trimmed or padded to two seconds. Samples were distributed into train, validation, and testing splits (80/10/10), ensuring an equal balance of acoustic and electronic samples and that samples originating from the same sample pack (i.e., recorded on the same drum in the same room) were contained within a single split.

We configure the sinusoidal tracker with a maximum number of sinusoids $N_S$=64, and a hop size of 256 samples. The noise generator uses $N_N$=128 noise bands and a hop size $h$=128 samples. We use a transient conditioning vector of length 128: $z \in \mathbb{R}^{128}$ and the TCN is composed of 8 blocks with a dilation factor of 2 and 32 hidden channels. While the TCN can process monophonic audio signals of any duration, all samples used in these experiments had a duration of two seconds. All encoders were optimized using a multi-resolution spectral loss with the same configuration of FFT, window, and hop sizes as in \cite{yamamoto_parallel_2020}, which has also been used in prior work using TCNs for audio effect modeling \cite{steinmetz_efficient_2022}. 
This loss is calculated as a weighted sum of the spectral convergence and log-magnitude spectral difference. We used an Adam optimizer with an initial learning rate of $1e^{-4}$ and a batch size of 12. Similar to \cite{steinmetz_efficient_2022}, the learning rate was scheduled to decrease by a factor of two if the validation loss did not decrease for 20 epochs, and training was halted if validation loss failed to improve for a further 20 epochs. All training runs were capped at 48 hours.

\subsection{Reconstruction Evaluation}
We consider three metrics to evaluate audio reconstruction: multi-resolution STFT error (same as the training loss, denoted MSS), log spectral distance (LSD) using the same formulation as \cite{birnbaum_temporal_2019}, and the mean absolute error between the spectral flux onset signals (SF) \cite{bello_tutorial_2005} extracted from $y$ and $\hat{y}$. The spectral flux onset signal is calculated as the $L_2$ norm on the rectified spectral difference:

\begin{equation}
    SF(n) = \sum_{\omega=0}^{\frac{N}{2} + 1}H(|X_{\omega}(n)| - |X_{\omega}(n-1)|)^2
\end{equation}

\noindent where $H(x)$ is a rectified linear unit, $H(x) = (x + |x|) / 2$, which includes only positive differences to emphasize the onset. Because applying a log transformation de-emphasizes transients, the LSD provides insight into the reconstruction of the signal with the decay emphasized, whereas SF error provides insight into the reconstruction of signal transients and onsets.

Reconstruction results computed on acoustic and electronic percussion samples, separately and combined, are shown in Table \ref{tab:overall-metrics}. Additionally, reconstruction metrics are shown for a selection of individual percussion instruments: kicks, snares, toms, and cymbals (which includes hihat) in Table \ref{tab:instrument-metrics}. %
Looking at the spectral flux onset error, the transient TCN models achieved the lowest error in all tests except for the individual cymbal sounds, where the sines plus noise model outperforms TCN models. The sines plus noise model achieved the best reconstruction results in terms of LSD for all tests. Interestingly, the MSS loss, which was the training objective, was lowest for transient TCN models, except for the individual cymbals. Investigating the separate spectral loss terms contributing to the MSS (i.e., spectral convergence and log-magnitude spectral difference), we note that models containing a TCN more efficiently decreased the spectral convergence term compared to the log-magnitude spectral difference. %

These results point to the following four takeaways: 1) Using a transient TCN generally improved signal onset reconstruction for membranophones (i.e., kick, snare, tom); 2) Sinusoidal plus noise without the TCN worked the best for idiophones (i.e., cymbals and hihats); 3) Generally, models that use noise summed in parallel performed better than models without; and 4) The transient TCN more effectively reduced spectral convergence loss during training, which appears to be correlated with improved onset reconstruction. The implication is that TCNs improved onsets at the expense of the signal decay. This was verified in an informal listening test during which we noted the addition of artifacts, similar to comb filtering, in the decay portion of TCN reconstructions.

\subsection{Visualizing the Transient Embedding}
\begin{figure*}[ht]
    \centering
    \includegraphics[width=1.0\textwidth]{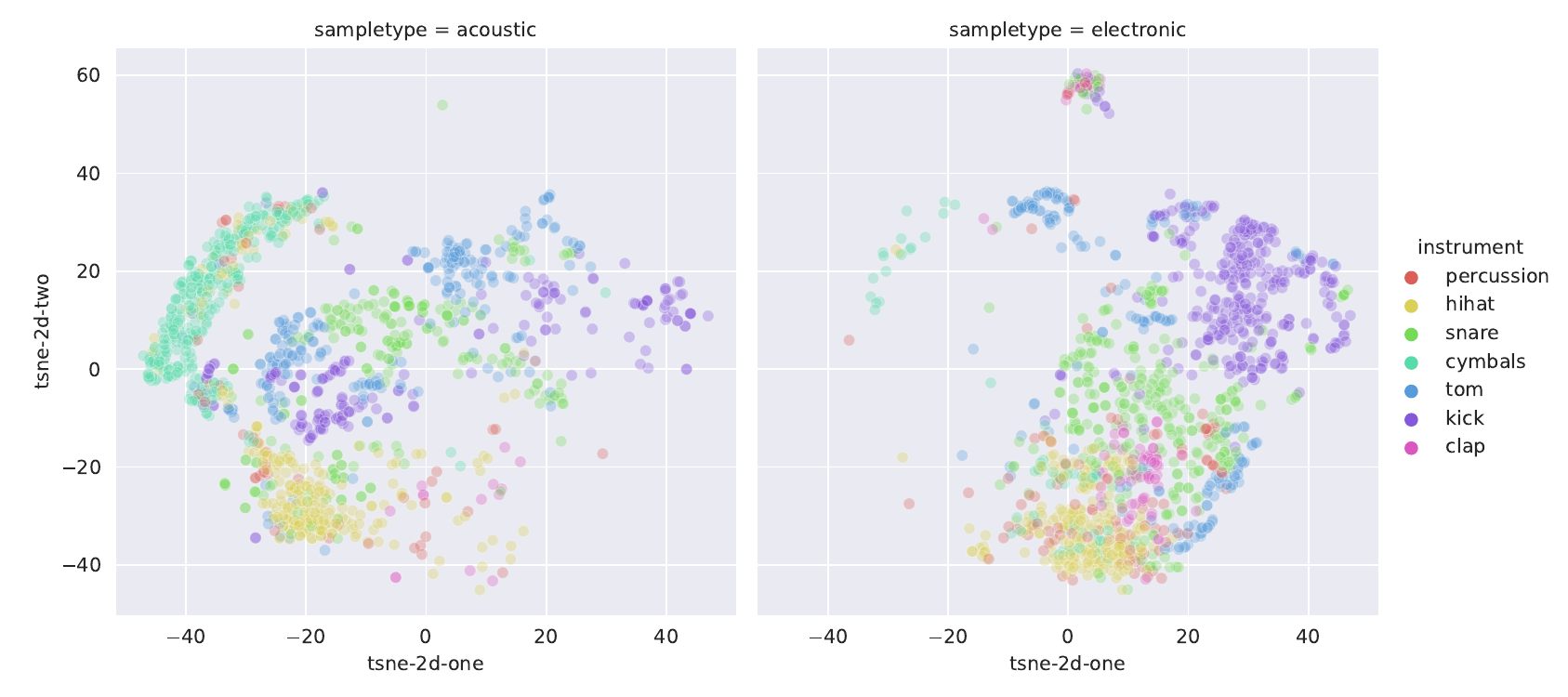}
    \caption{t-SNE plots of transient embeddings $z$ colored by percusison type. Left plot is acoustic instruments and right plot is electronic.}
    \label{fig:tsne_instruments}
\end{figure*}

We visualized the test dataset by producing embeddings using the trained transient encoder to provide further insight into the transient parameter space. For each audio sample, the result is an embedding vector $z \in \mathbb{R}^{128}$. For visualization, we use t-SNE \cite{van_der_maaten_visualizing_2008}, a common dimensionality reduction technique used to visualize high-dimensional data, to map embeddings to a two-dimensional space. Figure \ref{fig:tsne_instruments} shows the resulting 2D mappings; samples are separately plotted for acoustic and electronic samples and are colour-coded by instrument type to highlight clusters based on instruments within the embedding space. 
The embedding space features shared characteristics within instrument and sample types, showing small overlaps across these groups.
This indicates that there is a continuum in the space, where in many cases, different instruments in a vicinity feature similar embeddings. However, there is a progressive variation across instrument and sample types, with transient characteristics becoming more salient at specific locations. For instance, the electronic and acoustic sounds clearly occupy separate areas of the point cloud. We additionally note the relatively large distance between  kicks, hihats, and cymbals groups.

Observing the samples in the point cloud, we can infer that the embedding of our encoder represents well the sound characteristics required by the different samples. This opens up the possibility to parameterize the space to gain high-level control over the transients of drum sounds. This could serve as a first stepping stone towards controllable differentiable drum synthesis.

\section{Conclusions}
We presented a neural audio synthesis architecture for drum modeling in terms of three major sound components. In particular, we used a TCN modulated with FiLM embeddings as a method to reconstruct transient / onset signals within a sines plus noise synthesis model. We trained parameter encoders for transient and noise signal generators on a diverse dataset of acoustic and electronic drumset sounds and evaluated the resulting models using reconstruction metrics. Although we modelled transients indirectly using MSS loss, %
the TCN learned to model short-term signal features associated with membranophone onsets. We note that this came at the expense of added artifacts during the decay portion of the signal, which is reflected by the LSD. This points to the need for further work looking at how we can balance resynthesis fidelity of transient and decay signal components.

Our work takes a step towards a fully differentiable model for percussive synthesis using a neural source-filter approach for transient modeling. One major technical roadblock to realizing a fully differentiable percussion synthesizer is frequency estimation using gradient descent. Future work may leverage recent findings by Hayes et al. \cite{hayes_sinusoidal_2023} for this problem.
Beyond exploring solutions to these technical challenges, future work includes investigating the affordances of our analysis-synthesis framework for creative practices. For instance, by leveraging the latent space of a variational autoencoder for high-level control of each sonic component.

\section{Acknowledgments}
The authors would like to extend their gratitude to Ben Hayes for his contribution to a codebase that supported this research and for the helpful discussions on DDSP. Thank you to the two anonymous reviewers and to Lewis Wolstanholme for their helpful comments which improved the quality of this paper. This work is supported by the EPSRC UKRI Centre for Doctoral Training in Artificial Intelligence and Music (EP/S022694/1).
This research utilised Queen Mary’s Apocrita HPC facility, supported by QMUL Research-IT. \url{http://doi.org/10.5281/zenodo.438045}.

\bibliography{references}

\end{document}